# Self-Expression and Sharing around Chronic Illness on TikTok

**Rachael F. Zehrung, MS,[1] Yunan Chen, PhD[1]**
[1]University of California, Irvine, Irvine, CA, USA

**Abstract**
*While prior work has investigated the benefits of online health communities and general-purpose social media used for health-related purposes, little work examines the use of TikTok, an emerging social media platform with a substantial user base. The platform's multimodal capabilities foster creative self-expression, while the content-driven network allows users to reach new audiences beyond their personal connections. To investigate users' challenges and motivations, we analyzed 160 TikTok videos that center on users' firsthand experiences living with chronic illness. We found that users struggled with a loss of normalcy and stigmatization in daily life. To contend with these challenges, they publicly shared their experiences to raise awareness, seek support from peers, and normalize chronic illness experiences. Based on our findings, we discuss the modalities of TikTok that facilitate self-expression around stigmatized topics and provide implications for the design of online health communities that better support adolescents and young adults.*

**Introduction**
Peer support is crucial for health and disease management, as it provides patients and caregivers with informational, emotional, and social support beyond clinical encounters.[1–3] With the ubiquitous use of personal computers, such as mobile phones and laptops, healthcare consumers increasingly engage in online health communities (OHCs) where they seek health-related information and social support from others who may share similar health experiences or concerns. These communities manifest in different formats: While some are specifically designed for health (e.g., an asthma support group hosted by the Asthma and Allergy Foundation of America[4]), others are designed for general purposes but adapted by users for health concerns (e.g., a fertility group self-organized by users on Reddit[5]).

In this study, we investigated the lived experiences, challenges, and motivations of individuals with chronic illnesses. We chose TikTok, an emerging social media platform, due to its substantial user base and increased prevalence. Launched in 2017, TikTok is a short-form mobile video platform that allows users to create, share, and discover videos up to three minutes in length. Globally, around 70% of TikTok users are under the age of 35,[6] and it is one of the most used social media platforms among adolescents aged 13-17 in the United States.[7] It also appears that TikTok has become a place to discuss health, as evidenced by the 7.8 billion views of the tag #chronicillness.[8]

Compared to text-based platforms, TikTok provides additional modalities that can support self-expression, similar to how the visual modality of Snapchat has been found to enhance self-expression and sharing over text alone.[9] These modalities include video clips, photos, text, music, voiceovers, and animations, which users can combine to creatively express themselves and discuss challenging topics in an engaging manner. The platform is also content-driven and algorithmically curated to a user's interests, meaning that relevant video content is prioritized over content posted by close social connections as is the case with Facebook or Instagram, allowing users to reach broader audiences.

Although online communities have long been studied in health informatics literature, most prior work focuses on adult patients and text-based support,[10–14] while less work has investigated the experiences of adolescents and young adults (AYAs). Adolescence and young adulthood, defined as ages 13 to 25, are developmentally critical periods in which youth develop more independence, explore their identities, and expand their social networks. To better support young people in their independent health and disease management, it is crucial to understand their naturally adopted peer support practices in online communities such as TikTok, which is popular among youth. Such understanding could inform the design of OHCs that promote youth engagement and expression.

This study examines the online sharing practices of individuals with chronic illnesses on TikTok. Specifically, we investigate the following research questions: 1) What do users with chronic illnesses share on TikTok and 2) How do the modalities of TikTok facilitate users' self-expression and sharing?

To answer these questions, we conducted a content analysis of 160 TikTok videos that centered on users' firsthand experiences with chronic diseases. We found that TikTok users shared their experiences and challenges of living with a chronic disease, including a loss of normalcy and stigmatization in everyday life. In spite of these challenges, TikTok users shared their chronic disease content online to raise awareness, educate others, and find support from peers with shared experiences. Users engaged in these practices by using formats such as roleplaying and combining different modalities, providing new avenues for self-expression beyond what text-based online health communities currently offer. Based on our findings, we discuss how online health communities can better support self-expression around stigmatized topics and promote connectedness, particularly for AYAs.

**Methods**

To investigate the experiences, challenges, and support-seeking of people with chronic illnesses, we conducted a qualitative study to analyze TikTok videos that center on chronic illness experiences.

*Data Collection*

We scraped the data using Pyktok,† an open-source tool to collect video, text, and metadata from TikTok. Using the search terms "chronic illness," we collected 1,000 entries, which was the scraper limit for a single query, in November 2022. The results were screened based on the following criteria: (1) the video should be about personal experiences related to chronic illness and (2) the video should be in English. Since some users posted significantly more videos than others, to ensure that we captured the diverse experiences and perspectives of users, we included up to 5 randomly selected videos from the same user. We excluded videos that (1) were posted by other users (e.g., news platforms and organizations) and (2) were advertisements, because we are only interested in genuine, firsthand experiences with chronic conditions. We did not exclude any chronic conditions.

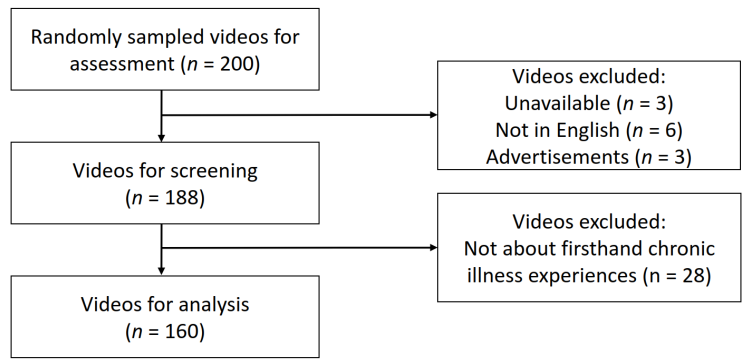

**Fig 1.** Flowchart describing the data set according to inclusion and exclusion criteria.

We randomly sampled 200 videos from the initial dataset of 1,000 videos, and then screened the sampled dataset according to the inclusion and exclusion criteria. RZ performed the screening process and consulted YC about specific videos in the event of uncertainty. As shown in Figure 1, we excluded videos that were unavailable at the time of analysis (3), not in English (6), were advertisements (3), and not about firsthand chronic illness experiences (28). We reached data saturation after 120 videos but analyzed an additional 40 videos to ensure full coverage.

*Data Analysis*

We performed a combination of deductive and inductive analysis. For the deductive analysis, we adapted Pretorius et al.'s[15] categorization of modalities used for TikTok videos and included the following categories that were relevant to the data: whether a video featured text overlay, the poster's own voice, use of music, roleplay, and text-to-speech (as shown in Table 1). We also manually recorded users' diagnoses if they were mentioned in the caption or video.

| Modality | Description | Percentage of posts (*n*=160) |
| --- | --- | --- |
| Text overlay | Text overlaid on the video, used for commentary or captioning | 76% (122) |
| Music | Music tracks, used in the background or for lip syncing and performing viral dances to capture viewers' attention | 70% (112) |

---

† https://github.com/dfreelon/pyktok

| | | |
|---|---|---|
| Poster's own voice | The user audibly speaking in the video | 42% (67) |
| Roleplay | The user acting out different characters (e.g., a discussion between themselves and a teacher) | 12% (19) |
| Text-to-speech | TikTok's text-to-speech capability, which reads text in a computerized voice. Used for comedic relief, accessibility, or when the user does not want to use their own voice | 6% (10) |
| Stitch | The user combines another TikTok video with their own. Often used to respond to open-ended questions or trends | 4% (6) |

**Table 1.** An overview of the formats, or stylistic elements, contained in the videos. A single video can incorporate multiple formats (e.g., a video can have music in the background and text)

Due to the short-form content style of TikTok videos, we were able to manually transcribe the videos. For the inductive analysis, we randomly selected 20 videos and individually performed open coding on 1) speech data: what users said in their videos; 2) textual data: captions and written content in the videos; and 3) visual data: photos and video clips contained in a single TikTok. We did not analyze the background music of the video in depth as it was beyond the scope of our analysis of users' chronic illness experiences, but we did note when music was used to participate in viral trends (e.g., dances). The authors met weekly throughout the coding process to engage in multiple rounds of discussion and review of the codes. In the event of discrepancies, we rewatched the videos and jointly came to a consensus. Then, we grouped conceptually similar codes into higher-level themes to structure our findings. For example, "get ready with me" and "nighttime feeding tube change" were merged into "routine." The first author individually coded the remaining 140 videos using the developed codes and additional codes when necessary.

Given that TikTok videos are considered publicly available data and our analysis did not contain any personally identifiable information, this study is considered IRB exempt in our institution. To further protect users' privacy, we used gender-neutral terms throughout the findings (they/ them/ theirs), referred to videos with identification numbers such as V1, and paraphrased quotes.

**Findings**

In this section, we provide an overview of the collected videos and then report on the types of content we identified. As shown in Table 2, the first main topic was users' experiences and challenges of living with a chronic disease, including dealing with the impact of illness on daily routine and encountering stigma regarding their diagnosis. The second topic relates to users' motivation for posting on TikTok, including a desire to raise awareness and relate to peers with shared experiences. Throughout the findings, we highlight how users leveraged TikTok's multimedia features and platform norms to augment their storytelling and communicate their experiences in an engaging way.

*Overview*

We observed that users actively posted on TikTok to share their experiences with chronic diseases. The 160 videos were collected from 122 unique vloggers. The videos ranged between 5 and 281 seconds, with a mean of 43 seconds and median of 31 seconds. Shorter videos allowed users to deliver content in an engaging way (e.g., answering common questions about their diagnosis while dancing) and share snippets of their everyday life, while (relatively) longer videos gave users space to discuss their challenges and journeys in more detail.

In terms of representation, the 122 vloggers show 44 unique chronic diseases. Users typically mentioned their diagnosis in the video or in the description; however, 55 videos did not provide this information. We refer to specific conditions in the context of users' experiences, but some examples of conditions represented in the data set include diabetes, postural orthostatic tachycardia syndrome (POTS), and inflammatory bowel disease (IBD).

| Topic | Theme | Description | Percentage of posts (*n*=160) |
|---|---|---|---|
| Experiences and challenges | Loss of normalcy | The video compares the user's life before and after diagnosis and/ or highlights the negative impact of illness on daily life | 13% (21) |
| | Routine | The video depicts everyday routines with chronic illnesses, but does not focus on negative aspects | 44% (70) |
| | Stigma and Stereotyping | The video focuses on users' experiences of stigmatization (e.g., being accused of faking their illness due to their appearance) | 21% (24) |
| Motivation to share on TikTok | Raising awareness | The video educates viewers or draws attention to the experiences of individuals with chronic illnesses (e.g., dismantling stereotypes) | 60% (96) |
| | Building community | The user shares or seeks advice and support. These videos are oriented towards other people with chronic illnesses rather than the general public. | 39% (63) |

**Table 2.** An overview of the types of content posted by users. A single video can address multiple themes (e.g., a user who discusses their experiences with stigma and then raises awareness about their condition).

*Users' Lived Experiences and Challenges Living with a Chronic Illness*
Videos centered on users' overall experiences and specific challenges living with chronic conditions, such as encountering stigma in everyday life. Despite these challenges, users shared their experiences on TikTok to raise awareness and educate healthy people about how to be better allies, as well as build community among other individuals with chronic illnesses and disabilities.

Loss of Normalcy: Diagnosis with a chronic illness can significantly impact all areas of life, from professional to personal settings. Users reflected on the loss of normalcy experienced after their diagnosis by showcasing the impact of illness on their daily routines and social interactions with others, such as teachers. They did not shy away from sharing their discomfort or frustrations, providing an authentic[16] look into their challenges living with chronic illness. For example, V86 shared, "*I used to love to smile… until I got Bell's palsy*," accompanied by contrasting videos of them laughing and smiling before diagnosis and then struggling to smile afterwards. By sharing video artifacts of before and after diagnosis, this user communicated the loss of normalcy and new sense of self-consciousness that they experienced around an everyday action (the ability to smile). Similarly, V61 compared their life before and after being diagnosed with a type of Ehler's Danlos Syndrome (EDS), a group of connective tissue disorders. They juxtaposed a video of themselves skiing before diagnosis with a video of themselves crying, overlaid with the text "*I wish I was dead.*" Through this contrast, the user candidly showed their despair over not being able to perform the activities that they previously enjoyed. Losing the ability to participate in their usual hobbies and social circles can cause patients to feel isolated, making social media an important tool for seeking support and finding community.

While V86 and V61 highlighted their loss of ability, other users emphasized the negative impact of their illness on everyday social interactions, particularly in school settings. V67 (Type 1 Diabetes) performed a roleplay in which they reenacted a scenario where the teacher mistook their medical device for a cell phone and reprimanded them in front of the class. Although the teacher in this scenario apologized after realizing their error, V67 still had to "*[try] not to cry*" because they "*hate being yelled at.*" Peer interaction and acceptance are particularly important during adolescence and emerging adulthood. By drawing attention to the user and her illness, the teacher in this scenario singled out the user and "othered" them from their healthy peers. Loss of normalcy is not only about the present or everyday activities; for some users, it also meant that their future plans from when they were healthy could no longer be realized in the way that they imagined. For instance, V38 recorded a video of themselves with a downtrodden expression, overlaid with the text "*becoming disabled at the age of 22 and no longer able to achieve any of my dreams.*" In emerging adulthood, the normative expectation is to be healthy, go to school or work, and make progress towards living independently. Yet, patients with chronic illnesses are set apart from these norms, and their trajectories are permanently altered. V75 vented a similar frustration and showed a video of themselves crying upon "*realizing that my chronic illness has no cure and this is my body forever.*" The combination of being diagnosed at a relatively young age and the permanence of chronic conditions presents a challenge for AYAs, who have to find a new sense of normalcy and cope with their symptoms. Through sharing their experiences on TikTok, users were able to reflect on the loss of normalcy and the broader impact of their condition on emotional well-being.

Routine: In contrast to users who highlighted the negative impact of illness on daily life, many users shared aspects of their everyday routines in a neutral manner to normalize the chronic illness experience. V2, for example, invited viewers to "*make dinner with me*" and showed how they set up their feeding tube. Likewise, V4 shared their "*tubie night routine*" which involved setting up their feeding tube and taking their nightly medications. These videos do not necessarily discuss users' challenges or the loss of normalcy. Instead, they invite viewers to be a part of the users' everyday lives with chronic illness, establishing "*sick normal*" (V111) as a common experience that evokes neither pity nor concern. By providing viewers with access to these private and often mundane moments, users were able to communicate the full extent of the chronic illness experience, as opposed to only the highs and lows. Some users even recorded active attacks and how they responded to them, highlighting that these scenarios are parts of their daily routine. For example, V52 recorded a "*get ready with me*" (GRWM) style video to show their daily makeup application and their fainting spells, stating:

> "*The reason I like to show my fainting spells is because I probably look fine to most people. Like right now, for instance, internally my body is screaming at me. There's quite a lot that can be going on behind the scenes.*"

V52 used the video as an opportunity to draw attention to invisible illnesses: Even though someone might "*look fine,*" they might experience other challenges "*behind the scenes.*" Throughout the video, the user fainted several times while applying their makeup, giving viewers an intimate understanding of how chronic illness impacts everyday routines.

Stigma and Stereotyping: Many users encountered disbelief or stigma regarding their illness, particularly in the case of invisible illnesses. By sharing anecdotes of other people assessing the authenticity of their diagnosis, users created a space to affirm their illness experiences and correct common misconceptions. For example, V33 roleplayed a conversation in which a "*Karen,*" or entitled woman, accused them of "*shoot[ing] up*" drugs and lying about their diagnosis ("*you don't have diabetes; you're not fat*"). In response to the "*Karen,*" the user explained the critical and lifesaving purpose of insulin. By performing this roleplay, V33 was able to vent about the judgment they encountered while also addressing misunderstandings about their condition. Common chronic illnesses such as diabetes may be well-known to the public, and as a result, people may have stereotyped expectations about how individuals with those conditions should look or act (e.g., the belief that diabetics are always "*fat*").

Similarly, V22 (Pernicious Anemia) expressed concern over stigmatization as an ambulatory wheelchair user. They recorded the process of exchanging their electric wheelchair battery during an outing and narrated:

> "*I need to stand up. Now I know that's quite a horrifying concept to people, where often if you stand up– the disabled person standing, it's a miracle! Or they were faking the entire time… that person was never disabled ever… I'm gonna stand up, which is very embarrassing when everyone can see me.*"

As this user suggested, wheelchair users might be expected to have no movement in their legs or be completely unable to walk, while in reality, some wheelchair users are ambulatory and may not use their wheelchairs all the time. Defying these stereotypes can lead to embarrassment or discomfort, as the user may be confronted and accused of "*faking*" their illness by bystanders. Likewise, V1 described their experiences as an ambulatory wheelchair user through a roleplay scenario:

> "*We were going for a walk. There was an inaccessible step, and since I was energetic and able, I walked up the steps myself. You said, 'Wait, you can walk?' and I said yeah, in short distances. You said, 'so you don't actually need a wheelchair, you're just lazy' … your whole worldview collapses the moment it's an ambulatory wheelchair user. Everyone is pro-accessibility until it doesn't fit their idea of disabled.*"

Many chronic illnesses can cause fatigue or weakness, which can explain why a wheelchair might be necessary even though an individual can walk short distances. Both V22 and V1 identified as chronically ill and disabled, positioning themselves at the intersection of two communities and highlighting the unique challenges that they experience. These shared experiences show that stigma and stereotyping are not individual occurrences, and that they are instead a byproduct of society's "*idea of disabled*" and chronically ill individuals.

Just as wheelchair users shared encountering doubt and judgment when they could walk, users with invisible illnesses faced additional skepticism around their diagnosis because they appeared to be typically healthy young adults. V119, an individual with Fibromyalgia and Lupus, shared their frustration in tears: "*I look the way I look, and people are like 'oh, what? I would never know' and I'm like, of course you wouldn't! Because I don't want to live my life looking the way I feel.*" This user recorded themselves in a vulnerable moment (i.e., crying in the car), an example of how users turn to TikTok as a place to share authentic experiences and emotions in the moment. Alternatively, V151 took a more playful approach by roleplaying a conversation between themselves and "*some random lady*" who confronted them about using priority seating on the bus and told the user, "*well, you look fine to me.*" To enhance this roleplay, the user employed a greenscreen to set the scene and make it clear that the scenario occurred on a public bus. By leveraging the platform's multimedia and special effects features to augment their storytelling, users like V151 were able to share their experiences in an engaging way.

While V119 indicated that she tried to look better than she felt, another user shared an alternative strategy for how to have other people "*take you seriously*": "*You can't look too good. Don't put on any makeup. Otherwise, they might think you're faking your illness*" (V106). A consistent concern expressed by users was being accused of "*faking*" their diagnosis, placing them in a position where they felt compelled to prove their illness or modify their behavior to meet expectations. As a whole, users engaged in a difficult balancing act to manage their self-presentation as both young and chronically ill, navigating social pressures and judgment in everyday life.

*Motivation to Share on TikTok*
Despite experiencing a loss of normalcy and encountering stigma in everyday life, we found that young adults disclosed their experiences on TikTok to 1) raise awareness and educate others and 2) build a sense of community that offers practical and emotional support.

Raising Awareness: Users turned to TikTok to raise awareness about the overlooked challenges and everyday experiences faced by people with chronic illnesses, with the goal of normalizing chronic illness and teaching people how to be better allies. While users drew upon their lived experiences, they did so with the purpose of educating healthy people, as opposed to posting for their personal benefit (e.g., venting). V104, an individual with Lyme disease, summarized their experiences through a roleplay scenario in which they acted as themselves and their co-worker to show "*what having a chronic illness is like.*" In the roleplay, the co-worker comments on how the user is "*sick all the time,*" gives unsolicited advice on "*how to stay healthy,*" and questions "*if it's all in your head… like you're just doing it for attention.*" This user was able to communicate more through a brief roleplay than a lengthy text description, giving them the opportunity to touch upon multiple challenges that people with chronic illnesses face in everyday life. By acting out a co-worker's negative or unhelpful behavior, V104 also demonstrated to viewers an example of what *not* to do when interacting with someone who has a chronic illness. Likewise, users dismantled stereotypes around people with specific chronic illnesses and sought to educate users about the negative impact of stereotyping. V136 features two users who created a song about their respective experiences with bipolar and diabetes. They employed a range of tactics to capture the audience's attention throughout the video, including dancing and dressing in different costumes. Directly addressing both "*the media [that] perpetuates stigma*" and viewers, the users implored "*don't call me crazy or say I'm lazy.*"

Users sought to raise awareness not only about their personal challenges, but the experiences and needs of people with chronic illnesses in general. As V111 explained,

> "*Sick and healthy people have completely different perceptions of things like pain and exhaustion… [a healthy person's] exhaustion is my baseline. So keep that in mind whenever a chronically ill person is talking about how tired they are or how much pain they're in, because generally they're always tired and they're always in pain, and if they're complaining about it, it's bad… healthy normal is not the same as sick normal*"

Through this explanation, this user highlighted that there is a different normative baseline for people with chronic illnesses and encouraged typically healthy people to be empathetic to that reality. Although everyone experiences some amount of pain or exhaustion, people with chronic illnesses have "*completely different perceptions*" because they are not in the same condition as people without illness. Another user (V128) illustrated this discrepancy by acting out a normal day in their life: "*POV: In a world where your energy level is shown above your head, most people start each day at 100%. You're a little different.*" Here, V128 asked viewers to imagine the portrayed experience from the user's point of view (POV) to help them understand life with chronic illness. The user showed that they start the day at 50% with an annotation overlaid on the video, and with each activity they completed, their energy level visibly decreased. Even enjoyable activities such as meeting with friends cost around 15% of the user's energy level, and by the end of the day, the user was left at 1%. Through the use of multiple content elements (text, music, and annotations), this video succinctly demonstrated everyday life as a chronically ill person and how different that experience is from "*most people.*" In summary, one of users' main motivations was to raise awareness and educate others about the chronic illness experience to foster more empathy.

Building Community: Users turned to TikTok to give and seek support from others with chronic illness, creating a sense of community and belonging. AYAs, even those with common illnesses, might not know other young people in real life with the same condition as them. V119 (Fibromyalgia and Lupus) directly sought peer support, asking "*I just want to know, people who have this, what do you do? I don't have anybody to support me.*" By posting about their challenges and experiences, users looked for other people with the same condition and people with chronic conditions more generally for advice on how to cope with their symptoms and deal with other people in real life, as well as emotional support. Some users also posted to vent about their negative experiences and seek encouragement from others. V112, an individual with sickle cell anemia, recorded themselves crying in the hospital because they did not receive adequate pain relief, stating "*you guys, I'm hurting so bad.*" While users sometimes posted asking for specific feedback, many users also turned to the platform to vent their frustrations and raw emotions, seeking emotional relief and support.

In addition to seeking support, some users created videos to provide support and knowledge to others with chronic illnesses. For example, V130 shared their techniques for pain management:

> "*Today we're going to talk about ways to relieve pain. My qualifications are that I have been in pain for 5 years and it's not going away anytime soon, but I've learned to cope with it... if I can help you discover some of these techniques sooner, that is all I ask for... I hope you feel better.*"

This user positioned themselves as an expert based on their own lived experiences with chronic pain and sought to leverage that expertise to teach pain management strategies to other people with chronic illnesses. V130 indicated that it took them a long time to learn effective techniques on their own and wanted to help others "*discover some of these techniques sooner*" to save them from the pain that they endured. Similarly, V125 shared one of their strategies to cope with chronic migraines after being unable to find adequate pain relief: "*This is an ice bowl of water. This is really good for migraine pain and all of that facial pain that you get.*" The user then demonstrated how they submerged their face in the bowl of water, staying there for longer periods of time with a snorkel. V125 captioned their video with several tags to reach out to other members of the community, including "*#spooniesoftiktok*" to reach people with chronic illnesses and "*#disabilitytiktok*" to reach a broader audience. Users sometimes engaged with communities for specific illnesses; V92, for example, stated "*welcome to the migraine side of tiktok*" and specifically welcomed other "*migraine friends.*" Another user with EDS also targeted their video towards an illness-specific community consisting of other people with hypermobility issues. V119 created a step-by-step educational tutorial to teach people how to create their own ring splints, explaining that "*a set like this normally sets you back $2000 so I made them myself. These cost $12. Here's how.*" The financial cost of treatment burdens some patients, and do-it-yourself solutions offered in a community forum can help users learn and practice effective self-management strategies.

Lastly, as opposed to illness-specific content, some users posted motivational or uplifting videos to inspire and support other people with chronic illnesses as a whole. As V50 shared:

> "*I have a couple of chronic conditions that make keeping the house clean really hard for me... I know I'm not the only one who struggles to keep things clean and stick to routines, so I just wanted to share whatever I have to give you motivation or make you know that you're not alone.*"

This user highlighted the challenge of having multiple chronic conditions and acknowledged that they were "*not the only one who struggles.*" By showing their difficulties and their attempts to overcome them, V50 sought to motivate others and let them know that they are not alone, and that they are instead part of a community with shared struggles and a support network.

**Discussion**

In this study, we investigated users' sharing and support-seeking practices on TikTok, which has high rates of adoption among individuals under the age of 35.[6] Based on our findings, we discuss how OHCs can better facilitate self-expression around stigmatized topics and promote patient engagement, particularly for adolescents and young adults (AYAs).

Providing Tools for Self-Expression: Prior work has found that OHCs can benefit patients with common chronic illnesses by providing informational and social support, though most work focuses on adult patients' experiences. AYAs with chronic conditions often struggle with feelings of isolation as they learn to manage their symptoms, and can benefit from meaningful social support.[17] Adolescent patients also want to be viewed as "regular" teenagers[18] and may conceal their diagnosis from peers on social media.[19] While OHCs targeted towards AYA patients might create a "safe space," existing OHCs are not well suited to the natural online behaviors and sharing practices of youth. Most OHCs are primarily text-based and centered around discussion forums, yet trends suggest that adolescents are turning away from text-based platforms to short-form visual content (e.g., TikTok and Snapchat).[7] Creating short videos could help patients convey their thoughts and emotions with less effort than writing, reducing the burden of seeking information and support.

These platforms also provide different modalities and methods of interaction compared to traditional OHCs, which users leverage to enrich their storytelling and discuss challenging experiences in an engaging, lighthearted way. For example, users employed green screens and filters to augment roleplaying scenarios by setting the scene (e.g., in the hospital) and creating distinct, engaging characters. By combining these media forms, users were able to share everything from snippets of daily life to form longer narratives while engaging viewers with text, music, animations, and special effects. Multimodal features that support creativity and self-expression might be particularly beneficial for the unique needs of this population by helping them normalize their feelings, experiences, and identities.[20] These visual features, if incorporated into OHCs, could help patients creatively represent their experiences and work through their feelings, similar to how people use blogs.[21] Prior work has highlighted the importance of aligning with the norms of existing platforms to facilitate personal data sharing.[22] Rather than implementing a new system, we suggest that OHCs leverage the existing norms and popularity of TikTok by encouraging patients to create videos on the platform and then share them with other members of OHCs. Integrating TikTok into existing OHCs might also introduce patients who are not as active on social media to new forms of self-expression in a familiar environment.

Fostering Patient Engagement: In addition to integrating multiple modalities for engagement and communication, users capitalized on TikTok's culture of virality and trends to raise awareness about chronic illness and engage with other AYAs with chronic illness. For instance, users created "stitched" videos to respond to trending questions from their perspective as individuals with chronic illness. They also incorporated trending audio clips and music to deliver engaging educational content, sometimes while performing viral dances or lip syncing. Users also adapted larger trends on TikTok and social media to a chronic illness context. For example, "Get Ready With Me" is a common format across social media platforms, in which influencers and celebrities walk through their daily routines by inviting viewers into their personal spaces. In this style of video, users demonstrated how chronic illness can make routine activities (e.g., getting dressed) more difficult, a display of "*sick normal*" as opposed to what is considered "normal" for healthy individuals. We observed that users often shared their raw emotions in a stream-of-consciousness fashion, and recorded videos in private spaces such as their cars, bedrooms, or hospital rooms, providing an intimate look into their "*behind the scenes*" challenges. By displaying both routine moments and moments of vulnerability, users opened windows into their authentic everyday experiences, with the goal of normalizing chronic illness experiences and encouraging healthy people to be more aware and understanding of people with chronic illnesses. Our findings echo a prior study on social norms of TikTok, where TikTok was considered a "fun" platform that normalizes expressions of both positive and negative emotions and experiences.[16]

Promoting Interest-based Sharing: Many social media platforms emphasize close social connections (e.g., showing content from friends and groups), while OHCs often group individuals by condition. TikTok offers a more dynamic approach to connectedness, as it prioritizes algorithmically curated content that matches the user's interests, regardless of their connection to the video creator. The viral and entertainment-driven nature of TikTok allowed users to appeal to diverse imagined audiences. These audiences included people with the same condition or other chronic illnesses (i.e., fellow "*spoonies*"), people with disabilities who had some experiential overlap (e.g., the use of mobility devices), and typically healthy individuals. Users aimed to share their experiences and advice to build a collective sense of identity and support other chronically ill individuals, as well as educate typically healthy people on how to be better allies (e.g., being empathetic towards people with invisible illnesses). OHCs might consider shifting from illness-specific groups and forums to affinity groups that connect patients with similar experiences, symptoms, or interests. For example, an affinity group might connect younger patients with invisible illnesses as they might encounter similar challenges in everyday life, particularly stigmatization.

Stereotyping and Stigmatization: We found tension in that users experienced instances of stereotyping in everyday life yet wanted to share their diagnoses and experiences on a public platform to raise awareness and build community with their peers. Due to the prevalence of common chronic illnesses, members of the general public might have their own set of expectations and ideas about individuals with those illnesses. For instance, as we noted in the results, users with diabetes recounted being stereotyped as "*lazy*" and "*fat*," as well as criticized for snacking to control their blood sugar levels. Stigma has long been studied in the context of common chronic illnesses[23,24] due to the negative impact of stigmatization (both experienced and anticipated) on care access and quality of life.[25] For AYAs with chronic illness, potential stigmatization and discrimination can also serve as barriers to disclosure with peers.[26] Yet, we observed that many users openly discussed their conditions and being negatively stereotyped in school, the workplace, and in public. While the public might be aware of common chronic conditions, this awareness does not necessarily translate into understanding. That is, individuals with common chronic conditions might be subject to even more stereotyping because of the pervasiveness of the condition. To refute common misconceptions about chronic illness, users shared their experiences of stereotyping and raised awareness about the diversity of chronic illness experiences and the overlooked challenges that chronically ill individuals face. Future work might examine comments on users' videos to explore practices around social support and online disclosure, as well as follow-up interviews with creators to understand their deeper motivations for posting about their experiences. Additionally, we acknowledge that there is a self-selection bias in that users opted to disclose their conditions and openly discuss their experiences. Future work should consider engaging with AYA patients who have not disclosed their conditions on social media, as they might have different perspectives around stigmatization.

*Limitations*

We were only able to study user motivations as manifested through their TikTok videos, which are short snippets of daily life. Follow-up studies might consider other methods to engage with participants directly, such as interviews. Furthermore, while we observed that the majority of users appeared to be AYAs, we could not verify users' ages and therefore did not use age as a screening criterion. Another known challenge of studying TikTok videos is the lack of transparency around algorithmic curation and its influence on search results. For example, new user profiles that execute the same search query may be shown different videos based on users' gender, age, and location.[27] Additionally, the tool that we used to scrape the data, Pyktok, comes with its own limitations. The tool only collected data after March 2019, and it is unclear from the documentation about how the videos were selected (e.g., most recent or most popular videos). We also acknowledge that our search strategy was not exhaustive. We used the search term "chronic illness," which might not encompass the full range of chronic conditions such as mental health conditions, eating disorders, and substance use disorders. Users might discuss these conditions with different terminology and as a result, they were not captured by the query. Furthermore, there is some amount of selection bias as TikTok users are only a portion of the population with chronic illnesses. Many individuals with chronic conditions are not on TikTok, and they might not disclose their conditions publicly or discuss their experiences in a similar manner. For these reasons, our data set might not be representative. Future research is needed to examine the breadth and depth of individuals' chronic illness experiences.

**Conclusion**

In this study, we analyzed TikTok videos to understand the lived experiences, challenges, and motivations of users with chronic illnesses. We found that users struggled with stigmatization and stereotyping in daily life, which motivated them to raise awareness about the diversity of chronic illness experiences and seek support online from peers with similar experiences. While some users emphasized the loss of normalcy experienced after diagnosis, we observed that they also sought to normalize chronic illness by showcasing their daily routines and inviting viewers into private moments or settings. Across all videos, users leveraged multimedia features on TikTok that facilitated their self-expression and allowed them to process stigmatized topics in a lighthearted and engaging manner. We hope that our findings can enrich future work that explores how to support engagement in online health communities through different forms of content creation and connectedness.

**Acknowledgements**

This work was supported in part by the National Science Foundation under Grant No. 2211923 and the Graduate Assistance in Areas of National Need (GAANN) program.